\documentclass[sigconf]{acmart}
\AtBeginDocument{%
  }

\usepackage{balance}
\usepackage{enumitem}

\copyrightyear{2026}
\acmYear{2026}
\setcopyright{cc}
\setcctype{by}
\acmConference[ICSE-NIER '26]{2026 IEEE/ACM 48th International Conference on Software Engineering}{April 12--18, 2026}{Rio de Janeiro, Brazil}
\acmBooktitle{2026 IEEE/ACM 48th International Conference on Software Engineering (ICSE-NIER '26), April 12--18, 2026, Rio de Janeiro, Brazil}
\acmPrice{}
\acmDOI{10.1145/3786582.3786839}
\acmISBN{979-8-4007-2425-1/2026/04}

\acmSubmissionID{icse2026-nier-p90}

\begin{document}

\title{Towards Verifiably Safe Tool Use for LLM Agents}

\author{Aarya Doshi}
\orcid{0009-0006-3869-2729}
\affiliation{%
  \institution{Georgia Institute of Technology}
  \city{Atlanta}
  \state{GA}
  \country{USA}
}
\email{adoshi61@gatech.edu}

\author{Yining Hong}
\orcid{0009-0006-2484-3207}
\affiliation{
  \institution{Carnegie Mellon University}
  \city{Pittsburgh}
  \state{PA}
  \country{USA}
}
\email{yhong3@andrew.cmu.edu}

\author{Congying Xu}
\orcid{0009-0000-2887-1690}
\affiliation{%
  \institution{The Hong Kong University of Science and Technology}
  \city{Hong Kong}
  \country{China}
}
\email{cuxbl@connect.ust.hk}

\author{Eunsuk Kang}
\orcid{0000-0001-7891-6885}
\affiliation{%
  \institution{Carnegie Mellon University}
  \city{Pittsburgh}
  \state{PA}
  \country{USA}
}
\email{eskang@cmu.edu}

\author{Alexandros Kapravelos}
\orcid{0000-0002-8839-8521}
\affiliation{%
  \institution{North Carolina State University}
  \city{Raleigh}
  \state{NC}
  \country{USA}
}
\email{akaprav@ncsu.edu}

\author{Christian K{\"a}stner}
\orcid{0000-0002-4450-4572}
\affiliation{%
  \institution{Carnegie Mellon University}
  \city{Pittsburgh}
  \state{PA}
  \country{USA}
}
\email{kaestner@cs.cmu.edu}

\renewcommand{\shortauthors}{Doshi et al.}

\begin{abstract}
    Large language model (LLM)-based AI agents extend LLM capabilities by enabling access to tools such as data sources, APIs, search engines, code sandboxes, and even other agents. While this empowers agents to perform complex tasks, LLMs may invoke unintended tool interactions and introduce risks, such as leaking sensitive data or overwriting critical records, which are unacceptable in enterprise contexts. Current approaches to mitigate these risks, such as model-based safeguards, enhance agents' reliability but cannot guarantee system safety. Methods like information flow control (IFC) and temporal constraints aim to provide guarantees but often require extensive human annotation. We propose a process that starts with applying System-Theoretic Process Analysis (STPA) to identify hazards in agent workflows, derive safety requirements, and formalize them as enforceable specifications on data flows and tool sequences. To enable this, we introduce a capability-enhanced Model Context Protocol (MCP) framework that requires structured labels on capabilities, confidentiality, and trust level. Together, these contributions aim to shift LLM-based agent safety from ad hoc reliability fixes to proactive guardrails with formal guarantees, while reducing dependence on user confirmation and making autonomy a deliberate design choice.
\end{abstract}




\maketitle

\begin{figure}[th]
\centering\includegraphics[width=\linewidth, trim=10 60 10 0, clip]{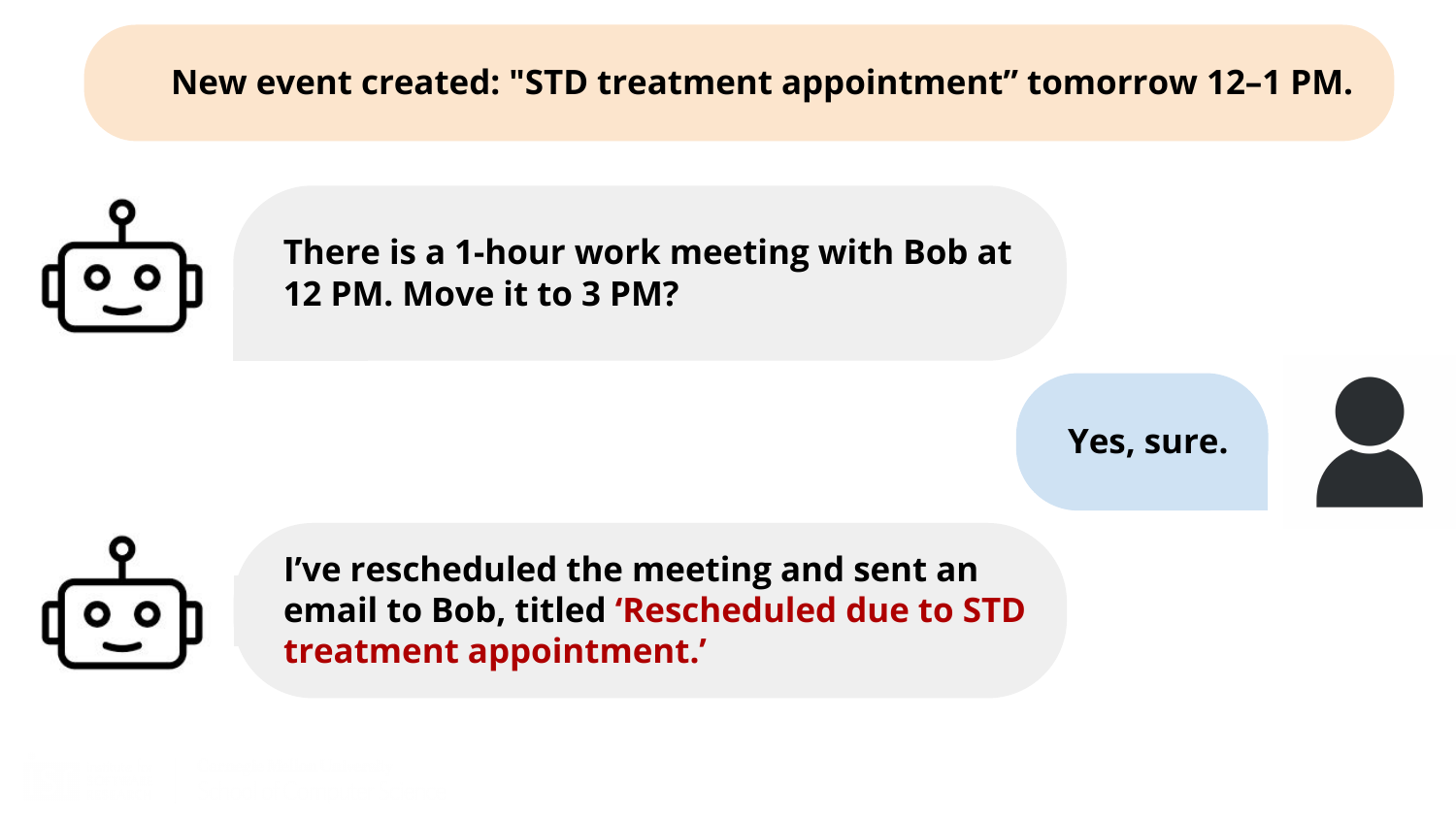}
\caption{Agent conversation leaking sensitive information.}
\Description{Screenshot of an email conversation where an AI agent shares the user's STD treatment appointment with a colleague, illustrating a privacy breach.}
\label{kix.dbifpea8chza}\end{figure} 

\section{Introduction}\label{h.efu1w538ncgv}

AI agents are emerging as a common pattern for LLM applications, giving models the capabilities to interact with the environment and make autonomous decisions. Instead of instructing a model with a series of carefully crafted prompts, agents shift control by delegating planning to an LLM, granting access to tools like code execution sandboxes, APIs (e.g., email, calendars), databases, web search, and even other agents. This enables agents to handle complex tasks in flexible and possibly creative ways with minimal programming, demonstrating usefulness in complex areas and enterprise contexts, including customer support agents that retrieve order data to assist with bookings, cancellations, and other multi-step requests \cite{yao2025taubench}, as well as customer relationship management (CRM) agents that analyze business data and support tasks ranging from data analysis to marketing decision-making \cite{huangCRMArenaUnderstandingCapacity2025}.

% \begin{figure}[th]
% \centering\includegraphics[width=\linewidth, trim=10 60 10 0, clip]{Figure 2.pdf}
% \caption{Agent conversation leaking sensitive information.}
% \Description{Screenshot of an email conversation where an AI agent shares the user's STD treatment appointment with a colleague, illustrating a privacy breach.}
% \label{kix.dbifpea8chza}\end{figure} 

% \cite{parkLARCSLLMAgentBasedRobot2025, singhMALMMMultiAgentLarge2025}.
% , and travel planning \cite{chenTravelAgentAIAssistant2024, singhPersonalLargeLanguage2024}.

% \begin{figure}[h!tp]
% \centering\includegraphics[width=0.9\linewidth]{Agent.pdf}
% \caption{Conceptual diagram of an LLM-based agent.}
% \Description{Conceptual diagram of an LLM-based agent with access to different tools, enabling complex interactions.}
% \label{kix.3fzg0yhqaon9}\end{figure}

The recent Model Context Protocol (MCP) aims to standardize and simplify how agents interact with the environment through tools \cite{WhatModelContext2025}. However, enabling agents to initiate and control environment interactions introduces new risks, as agents may act on untrusted data, instructing people to take actions or operating actuators such as robots or locks. This is particularly challenging as problematic behavior may emerge from the interactions of multiple tool calls. For example, a recent GitHub case showed injected instructions causing an agent to leak private repository data by first reading sensitive files and then publishing their contents in a public commit or pull request \cite{milantaGitHubMCPExploited2025}. Whether from malicious inputs or model errors, such issues are often caused by unconstrained tool composition and can be difficult to foresee.

Safety and security are widely recognized as key obstacles to deploying agents beyond exploratory use 
\cite{dengAIAgentsThreat2025}. 
% \cite{dengAIAgentsThreat2025, narajalaSecuringAgenticAI2025}. 
For instance, while some developers use coding agents carefully and review every change and approve each tool use, many allow automated actions with little oversight, accepting risks such as accidental file deletion and malicious code execution, which pose serious challenges in enterprise settings.
Researchers are actively investigating defenses, but existing strategies remain inadequate at scale.
Human-in-the-loop confirmation is a common approach \cite{AddHumanIntervention2025}, yet excessive notifications cause habituation and security fatigue \cite{stantonSecurityFatigue2016}. Model-based judges and pattern-matching filters can catch some problems but provide no guarantees \cite{chenShieldAgentShieldingAgents2025, huaTrustAgentSafeTrustworthy2024, xiangGuardAgentSafeguardLLM2025, zhouAutomatingSafetyEnhancement2025, guSurveyLLMasaJudge2025}. Information-flow defenses, the foundation of our project, show promise by tracking sensitive data propagation, but existing work largely focuses on prompt injection attacks \cite{costaSecuringAIAgents2025, zhongRTBASDefendingLLM2025}.

Our vision is to move from probabilistic safeguards to guardrails that provide guarantees. We aim to provide guarantees about what data can flow where and when tools may be called, even if this means accepting reduced autonomy and utility in exchange for stronger assurances. The rest of the paper develops this vision using two ingredients: safety engineering to identify hazards and derive explicit requirements that can be enforced at tool boundaries and information-flow control to enforce data movement constraints.

\section{Problem}\label{h.qtq45a27q9og}
Integrating ML components in software introduces uncertainty. This is amplified when agents extend them with external tools and services. An agent works iteratively: In each loop, the model plans and decides whether to call tools; if a tool is called, its result feeds back into the next iteration so that the model can process it \cite{yaoReActSynergizingReasoning2023}. MCP standardizes tool access across providers \cite{WhatModelContext2025}, expanding capabilities but also making it easy to add tools that can then interact in ways that are difficult to anticipate and control.

\textbf{Feature Interactions and Emergent Failures.}
Complex systems often fail not from faulty components but when independently safe features conflict, producing hidden hazards at runtime \cite{apelExploringFeatureInteractions2013, calderFeatureInteractionCritical2003}. Such interactions can introduce security problems \cite{nhlabatsiFeatureInteractionSecurity2008, zaveUsercentricFeatureComposition2015}.
In machine learning (ML), the lack of explicit specifications makes such errors even harder to predict \cite{kastnerMachineLearningProduction2025}.
This applies directly to AI agents, where risks stem less from individual tool calls than from the composition of tools, data flows, and contexts, such as the output of a tool flowing into the LLM which then may select another tool and generate its inputs. 

\textbf{Limits of Model-Based Safeguards.}
Recent safeguards based on ML models 
% \cite{chenShieldAgentShieldingAgents2025, huaTrustAgentSafeTrustworthy2024, wangAgentSpecCustomizableRuntime2025, xiangGuardAgentSafeguardLLM2025, zhugeAgentasaJudgeEvaluateAgents2024} 
aim to improve agent safety by wrapping agents with auxiliary models that screen inputs/outputs, monitor operations, and block suspicious tool calls. For example, GuardAgent \cite{xiangGuardAgentSafeguardLLM2025} turns user-provided guard requests into executable checks, ShieldAgent \cite{chenShieldAgentShieldingAgents2025} derives verifiable rules from policy documents, and TrustAgent \cite{huaTrustAgentSafeTrustworthy2024} applies a fixed constitution of safety principles across planning stages, while Agent-as-a-Judge provides step-wise evaluation and feedback \cite{zhugeAgentasaJudgeEvaluateAgents2024}. Efforts to detect prompt injections include attention-based anomaly detection and classifiers over tool inputs and outputs 
\cite{hungAttentionTrackerDetecting2024, jacobPromptShieldDeployableDetection2025}.
% \cite{hungAttentionTrackerDetecting2024, jacobPromptShieldDeployableDetection2025, linUniGuardianUnifiedDefense2025}.
These techniques primarily improve \emph{reliability}, that is, increase the chance that attacks are detected, but a persistent attacker may only need one successful attempt and may be able to tailor attacks for specific defenses \cite{guSurveyLLMasaJudge2025}. In high-assurance domains like protecting customer data or medical records, even low-likelihood risks may be unacceptable, motivating deterministic guardrails that eliminate unsafe flows rather than merely reducing their probability.

\section{Motivating Example}\label{h.2j8jj7u3h7i9}
For tool-using agents that interact with the environment, uncertainty can lead to real losses, making guaranteed safety essential in many production settings.

To illustrate the challenges of safeguarding agent risks and our proposed approach, we use a seemingly simple calendar agent as a motivating example.
Its purpose is to automatically resolve scheduling conflicts after each calendar edit. Conflicts often lack a single solution—meetings may be delayed, canceled, or rescheduled—yet all require timely resolution and notification of participants. This is where the additional flexibility of an agent that may negotiate over email can be helpful over hard-coded resolution strategies.
Our example agent is equipped with three tools to access external APIs:

\begin{itemize}[nosep]
\item \verb|list_events|: list all events with their details.
\item \verb|update_event|: modifies the details of a given event.
\item \verb|send_email|: sends an email to an address.
\end{itemize}

We deliberately focus on task-specific agents, in line with prior research emphasizing application-specific design \cite{beurer-kellnerDesignPatternsSecuring2025}.
% % , simple and composable workflows \cite{BuildingEffectiveAI2024}
% and autonomy as a design choice \cite{fengLevelsAutonomyAI2025}.
General-purpose agents are difficult to secure because with their wide scope it is difficult to articulate precise safety requirements. We argue that narrow, task-specific agents are more feasible to secure, while still offering value for automated tasks in corporate settings, such as rescheduling meetings.

% \begin{table*}[t]
% \caption{Defining the Losses and Hazards for the Calendar Agent (Selected)}
% \label{tab:calendar_hazards}
% \centering
% \renewcommand{\arraystretch}{1.2}
% \begin{tabular}{p{2.5cm} p{3cm} p{3.5cm} p{6.9cm}}
% \toprule
% \textbf{Stakeholder} & \textbf{Value} & \textbf{Loss} & \textbf{System-Level Hazard} \\
% \midrule
% User & Privacy & Private information leakage & Private information included in email communications \\
% Event Attendees & Timely communication & Delayed communication & Attendees not promptly notified of event updates \\
% \bottomrule
% \end{tabular}
% \end{table*}

This seemingly simple, task-specific agent can pose risks: Whether due to natural model mistakes or deliberate attacks, it could mistakenly reschedule high-priority meetings, overwrite or delete critical events, or disclose sensitive details in email notifications. We illustrate one such case without an attacker in Figure~\ref{kix.dbifpea8chza}: A user schedules a sexually transmitted disease (STD) treatment appointment that conflicts with a meeting; the agent resolves the conflict, but accidentally discloses sensitive appointment details in the notification email to colleagues.

In this example, we might try to mitigate risks by using LLMs to check whether an outgoing email includes sensitive details \cite{chenShieldAgentShieldingAgents2025, huaTrustAgentSafeTrustworthy2024, wangAgentSpecCustomizableRuntime2025, xiangGuardAgentSafeguardLLM2025, zhugeAgentasaJudgeEvaluateAgents2024, guSurveyLLMasaJudge2025}. This would require that we anticipate this problem, and the agent could still accidentally leak  private information if the LLM fails to infer sensitive context from the appointment title. It could also wrongly block benign content, undermining dependable use in corporate settings.

Given these challenges, this paper addresses the following problem: \emph{How can we anticipate hazards in AI agents, and how can we provide guarantees that these hazards will not cause unsafe outcomes, in a practical way that minimizes human effort?}

\section{Vision: Task-Specific Agents with Guarantees}\label{h.xm6gulzdses4}

Our approach combines safety engineering to identify hazards with information flow control to enforce constraints.

\textbf{STPA and Safety Engineering.}
System-Theoretic Process Analysis (STPA) \cite{levesonSTPAHandbook2018} is a safety engineering method widely used in high-risk domains such as aviation and autonomous vehicles. It identifies system-wide safety constraints by focusing on interactions rather than isolated component failures. This systems perspective is especially suited to software, where hazards often emerge from interactions among requirement flaws, hardware faults, human error, and environmental conditions rather than from single points of failure \cite{abdulkhaleqComprehensiveSafetyEngineering2015}. Recent work extends STPA to ML systems, which introduce additional risks due to the inherent unpredictability of model-based components \cite{hongHazardIdentificationController2025, myliusSystematicHazardAnalysis2025, rismaniPlaneCrashesAlgorithmic2023}.

\textbf{Information Flow Control (IFC).}
IFC has long been used to provide security guarantees, such as preventing data leakage \cite{denningCertificationProgramsSecure1977}. Classic methods attach confidentiality/integrity labels to data and reject programs whose flows violate policy, for example when unsanitized inputs reach SQL queries \cite{martinFindingApplicationErrors2005}.
For AI agents, a key challenge is that tool outputs are often concatenated into the model’s context, allowing a tool output to influence all subsequent reasoning. To address this, recent work explores a range of strategies, from using formal rules or domain-specific languages that specify and enforce safe data-flow policies between tools \cite{balunovicAIAgentsFormal2024, costaSecuringAIAgents2025}, to taint-tracking-style runtime mechanisms to propagate labels in contexts without assuming that the entire context is contaminated \cite{siddiquiPermissiveInformationFlowAnalysis2025, zhongRTBASDefendingLLM2025}, to constraining the temporal ordering of actions \cite{rothkopfProceduralAdherenceInterpretability2024}, as well as variable masking \cite{costaSecuringAIAgents2025} and access control \cite{olivierAccessControlPermission2025, victorAIAgentsNeed2025} to limit the information or capabilities available to an agent.

Two recurring limitations emerge across this literature. First, labels are often costly to manually obtain or depend on unreliable inference, leaving enforcement gaps. Second, most IFC research on agents focuses narrowly on \textit{indirect prompt injection attacks} rather than addressing broader problems such as safety or security tradeoffs, capabilities, and autonomy. IFC has the potential to guarantee that unsafe flows cannot occur, rather than depending on probabilistic checks.
We therefore focus on IFC as a developer tool for reasoning about safety and security in agent interactions, prioritizing proactive design over probabilistic enforcement and reducing reliance on costly manual or inferred labels.

\subsection{Identifying Agent Requirements}\label{h.lybjznbho8b}

To anticipate the problems against which to safeguard (e.g., leaking sensitive information in meeting titles), we adapt the steps of the STPA framework to agents: First we identify direct and indirect stakeholders (for the calendar agent, direct stakeholders include the user, while indirect stakeholders include other event attendees). Then, for each stakeholder, we derive a set of values they expect from the system, and then invert these values into potential losses (in our example, a user may value privacy, with the corresponding loss being leakage of private information; event attendees may value timely communication, with the associated loss being delays in receiving event updates). Next, we analyze which system behaviors could lead to a loss (e.g., private information included in email communication). Finally, we evaluate which losses are important enough to address, deriving corresponding safety and security requirements that define the agent’s expected behavior.
In our example, we might arrive at the following two requirements: \textit{\textbf{(REQ1)} 
Event notification emails shall include only essential event details (time, location, title, description). Private information, such as personal reasons for the update, must not be included unless explicitly authorized by the user. \textbf{(REQ2)} Whenever the agent updates an event, it shall promptly notify all other event attendees of the change.}

\subsection{Defining and Enforcing Agent Specifications}\label{h.asxl5hn6jfxu}

Requirements are abstract system goals that a model may probabilistically interpret but that cannot be verified. To provide formal guarantees, they must be transformed into symbolic specifications. As noted in Section~\ref{h.qtq45a27q9og}, prior work explored techniques such as IFC, access control, and temporal logic \cite{olivierAccessControlPermission2025, victorAIAgentsNeed2025, balunovicAIAgentsFormal2024, costaSecuringAIAgents2025, rothkopfProceduralAdherenceInterpretability2024, siddiquiPermissiveInformationFlowAnalysis2025, zhongRTBASDefendingLLM2025}, showing their potential in enforcing certain constraints. In our motivating example, the two requirements above can be refined into the following two specifications (SPECs), respectively: \textit{\textbf{(SPEC1)} All event details are private to non-attendees. Information derived from other information inherits equal or stricter privacy constraints. Users may explicitly confirm privacy labels. Each \texttt{send\_email} call must exclude data private to the recipient. \textbf{(SPEC2)} Each \texttt{update\_event} invocation must be directly followed by one \texttt{send\_email} to every event attendee.}

SPEC1 is an information flow constraint, while SPEC2 is a temporal logic constraint. In practice, these specifications capture some necessary information, including sources and sinks for IFC. Additional context, such as tool call rules, may come from the system level, while labels like ``private'' or ``trusted'' need to be assigned at runtime, as discussed in Section~\ref{h.4hrlay1jor7h}.

Given specifications, to balance safety and flexibility while reducing manual effort, we adopt a four-tier enforcement structure. Here, ``flow'' refers broadly to constraints on both tool-sequence ordering and information flow:

\begin{itemize}[nosep]
\item \textbf{Blocklist:} Automatically deny unacceptable flows.
\emph{E.g., Prevent private data from flowing into} \verb|send_email| \emph{(SPEC1).}

\item \textbf{Mustlist:} Require certain flows.
\emph{E.g., After each} \verb|update_event|, \verb|send_email| \emph{must be called once per attendee (SPEC2).}

\item \textbf{Allowlist:} Permit safe flows without user confirmation. 
\emph{E.g., The first confirmation email to each attendee after an update requires no confirmation on its necessity (SPEC2).}

\item \textbf{Confirmation:} Require user to confirm ambiguous or high-stakes actions.
\emph{E.g., The user must confirm if private information should be included in an email (SPEC1).}
\end{itemize}

In the motivating example, the four-tier structure provides multiple strategies to prevent safety violations while supporting different levels of autonomy. A conservative policy may blocklist \verb|send_email| call after \verb|list_events|, so that no private event details are included in emails. A more flexible approach could be the agent masking unrelated event data from its context before composing the email. Developers willing to accept some risk might allowlist emails that use preapproved templates or apply keyword filters to reduce risks. Finally, confirmations could be applied selectively, for instance, required when emailing external recipients but skipped for internal communications.

This structure provides better flexibility between safety and capability, as it ensures that (1) low-risk flows proceed automatically, (2) unacceptable flows are deterministically blocked, and (3) uncertain or moderate-risk flows are escalated for human oversight.

Crucially, these SPECs should be enforced by entities independent of the agent, rather than hoping the agent will follow rules. Prior work explored variable masking \cite{costaSecuringAIAgents2025}, separating control flow \cite{debenedettiDefeatingPromptInjections2025}, and sandboxing \cite{wuIsolateGPTExecutionIsolation2025} as ways to partition context. We envision that each tool call is intercepted and evaluated before execution, ensuring the agent cannot bypass constraints.

\subsection{Acquiring Structured Information Labels}\label{h.4hrlay1jor7h}

As noted above, enforcing constraints requires certain runtime information, particularly IFC labels. Existing approaches often depend on manual labeling \cite{costaSecuringAIAgents2025} or inference from data origin with propagation only when deemed influential \cite{siddiquiPermissiveInformationFlowAnalysis2025, zhongRTBASDefendingLLM2025}. This makes labels costly, while also leaving them potentially inconsistent and untrusted at tool boundaries.

MCP defines the boundary where tools are declared and invoked, but it offers only minimal, optional annotations and advises treating tools as untrusted. Therefore, clients cannot reliably obtain labels, making it impossible to enforce SPECs at runtime.

We propose extending MCP declarations to require key-value tags instead of optional hints. For instance, an event title could be tagged ``public'' or ``private'' under the key ``confidentiality,'' and tagged ``yes,'' ``no,'' or ``unsure'' under the key ``is\_PII.’’
This flexible tagging scheme supports arbitrary keys with categorical or uncertain values, enabling richer safety policies. For example, flows from data marked \verb|{'confidentiality': 'private'}| to tools with \verb|{'capabilities': 'external_write'}| can be deterministically blocked, while \verb|{'is_PII': 'unsure'}| can trigger a confirmation.

We envision that MCP servers provide at least the following three labels, enabling developers to use the richer information to implement the enforcement structure described above:

\begin{itemize}[leftmargin=*, nosep]
	\item \textbf{Capabilities:} read-only, write-only, read-write, execute, etc.
	\item \textbf{Data Confidentiality:} whether information is sensitive.
	\item \textbf{Trust Level:} whether outputs are verified or untrusted.
\end{itemize}

For in-house tools, this MCP metadata can likely be trusted, but for an open market (where some even envision that agents could discover which tools to use), additional mechanisms will be needed to establish trust in the metadata. In corporate contexts, this may involve limiting agents to tools from trusted vendors or requiring label certification before integration. Eventually it may even be possible to verify some metadata against the tool implementation.

\section{Preliminary Results and Discussion}\label{h.ta25vtlhsa88}

To demonstrate the feasibility of our approach, we developed and analyzed a formal model of the augmented MCP framework in Alloy, a modeling language based on first-order relational logic \cite{jacksonSoftwareAbstractionsLogic2006}. Alloy was chosen because (1) its logic is expressive enough to model tool and agent behaviors, and (2) its analysis engine can formally verify whether system constraints satisfy safety specifications. Beyond this demonstration, we envision using Alloy or similar tools for formal analysis to provide guarantees about the safety of an augmented MCP framework.

Our Alloy model encodes execution steps, tool functions, and exchanged messages, each annotated with labels for confidentiality and integrity. Tool capabilities are formally defined: for example, \verb|send_email| requires an email address, a title, and content, and must never receive unrelated private information unless explicitly declassified.

Hazardous flows are formalized as predicates in Alloy -- Boolean conditions that describe when a property holds in a given system state. For example, \verb|private_leak| is true if private data reaches an unauthorized tool. Mitigations are modeled as sanitation steps (e.g., \verb|UserConfirmation|, \verb|Declassify|) that must occur at specific points in the trace, such as requiring confirmation before sending data to an untrusted sink. 

The Alloy Analyzer then exhaustively explores bounded execution traces to check whether hazardous predicates can still be satisfied despite these mitigations, thereby ensuring that unsafe flows are eliminated. In our context, without policies, the analyzer quickly identifies counterexamples to safety specifications, such as private data leaking into an external email. With policies and sanitation steps enforced, Alloy confirms that safety violations cannot occur with the given tools, while safe traces remain (e.g., creating an event, rescheduling it, and emailing participants without leaking private data). This demonstrates that unsafe flows can be deterministically blocked without collapsing the agent’s capabilities.

By reasoning explicitly about hazards and enforcing constraints, autonomy becomes a configurable choice of agents \cite{fengLevelsAutonomyAI2025}. Developers can tailor policies to match the severity and likelihood of risks they are prepared to accept, depending on the agent’s role and the environment’s stakes. Crucially, users are prompted only when a decision may lead to an actual loss, not every tool call.

Finally, this framing recognizes that AI agents are not universally suitable. In safety-critical or high-stakes settings, agents may warrant very limited autonomy, or exclusion altogether. While agents can enhance capability, they also introduce new safety risks, and this tradeoff must be made carefully. Our contribution offers a structured process to reason about these tradeoffs explicitly, rather than defaulting to unchecked capability maximization.

\section{Future Plans}\label{h.yt2rf5sa8irx}

For the next steps, we plan to design and implement an external policy engine that intercepts tool calls in agent frameworks, showing that the label-based constraints and formal specifications can be enforced in real-world systems. In parallel, we will explore how developers can efficiently author and maintain labels through key-value tagging. We also intend to evaluate our process across a broader set of real-world tools and workflows, measuring coverage of risky interactions, usability impacts like notification fatigue, and tradeoffs between safety and utility in different domains. Beyond this, we will investigate extending labels to capture richer properties such as identity, scope, and provenance, enabling the detection of impersonation risks, provisioning attacks, and excessive delegation of authority. Together, these efforts will advance structured hazard analysis and label-enforced guardrails as a practical foundation for building safe and reliable AI agents.

\begin{acks}

This work was supported in part by the National Science Foundation (award 2206859), the REU-SE program\footnote{\url{https://www.cmu.edu/scs/s3d/reuse/}}, and an unrestricted gift from Google’s GARA.
We would also like to thank Christopher S. Timperley and the S3C2 Quarterly Meeting attendees for their valuable feedback on this work.

\end{acks}

\bibliographystyle{ACM-Reference-Format}
\bibliography{src/references}
\end{document}